\documentclass[prb,preprint,showpacs]{revtex4}
\usepackage{graphicx}
\usepackage{amsmath,amssymb,bm}
\begin{document}
\title{Model of ripples in graphene }
\author{L.L. Bonilla$^{1}$, A. Carpio$^{2}$ }
\affiliation {
$^1$G. Mill\'an Institute, Fluid Dynamics, Nanoscience and Industrial
Mathematics, Universidad Carlos III de Madrid, Avda.\ Universidad 30; E-28911 Legan\'es, Spain\\
$^2$Departmento de Matem\'atica Aplicada, Universidad Complutense de Madrid; E-28040 Madrid, Spain\\
}
\date{\today}
\begin{abstract}
 We propose a model of ripples in suspended graphene sheets based on plate equations that are made discrete with the periodicity of the honeycomb lattice and then periodized. In addition, the equation for the displacements with respect to the planar configuration contains a double-well site potential, a nonlinear friction and a multiplicative white noise term satisfying the fluctuation-dissipation theorem. The nonlinear friction terms agree with those proposed by Eichler et al [Nature Nanotech. {\bf 6}, 339 (2011)] to explain their experiments with a graphene resonator. The site double-well potential indicates that the carbon atoms at each lattice point have equal probability to move upward or downward off-plane. For the considered parameter values, the relaxation time due to friction is much larger than the periods of membrane vibrations and the noise is quite small. Then ripples with no preferred orientation appear as long-lived metastable states for any temperature. Numerical solutions confirm this picture. 
\end{abstract}
\maketitle 
\renewcommand{\thefootnote}{\arabic{footnote}}

\section{Introduction}
\label{sec:1}
The first visualizations of atoms in suspended graphene sheets showed that they were covered with ripples.\cite{mey07} These ripples are several nanometers long waves of the sheet without a preferred direction.\cite{mey07,ban09} The initial preparation of a suspended graphene bilayer may result in \AA ngstrom-sized ripples that are absent in graphene monolayers.\cite{mao11} In experiments on graphene sheets suspended on substrate trenches, there appear much longer and taller waves (close to a micron scale) directed parallel to the applied stress.\cite{bao09} These long wrinkles are thermally induced and can be explained by continuum elasticity.\cite{cer03}  Ripples are expected to be important for electronic transport in graphene,\cite{gei08} and there is active research about the effects of ripples and strain on electronic properties, including possible strain engineering.\cite{gui09,voz10}

The earliest theoretical studies of ripples using Monte Carlo\cite{fas07} or molecular dynamics simulations\cite{abe07} have shown that ripples may be connected to variable length $\sigma$ bonds of carbon atoms and may be due to thermal fluctuations. Other studies explored the connection between rippling and electronic properties\cite{kim08,gaz09} and suggested that, at zero temperature, the electron-phonon coupling may drive the graphene sheet into a quantum critical point characterized by the vanishing of the bending rigidity of the membrane.\cite{sgg11} There is a buckling transition between the regimes of positive and negative tension in graphene. Then the inherent tension in suspended graphene produces ripples with a characteristic length and aspect ratio.\cite{sgg11} Thus rippling already occurs at 0K and it is due to the strong interplay between the dynamics of the free electrons in the membrane and its mesoscopic structure. While most of rippling studies are focused in equilibrium behavior, in this paper we want to analyze the dynamics of ripples using a model based on discrete elasticity that incorporates nonlinear friction and white noise sources. 

When observing ripples with a high-resolution transmission electron microscope, the graphene sample is bombarded by a low-intensity electron beam. Even though the electron beam does not have sufficient energy to knock off carbon atoms, it may create defects by inducing bond rotation and certainly excite the atoms. Thus an observed graphene sample is continuously excited and cannot be considered to be in  equilibrium.\cite{mey07,ban09} We have suggested that the electron beam may push the carbon atoms vertically away from the planar configuration of the graphene sheet in a random fashion and its effect could be modeled by coupling each carbon atom to an Ising spin that tries to move it off-plane.\cite{bc12} In a simpler context, one-dimensional (1D) mechanical systems coupled to Ising spins that flip randomly according to Glauber dynamics exhibit ripples at any temperature in the limit as the spin relaxation time is much longer than the vibration periods of the mechanical system.\cite{bcpr11} In this spin-string system, the ripples are metastable quasi-equilibrium states about which the mechanical system experiences rapid oscillations. Both the 1D spin-string and the spin-membrane (graphene) models exhibit a buckling transition at a critical temperature, above which the string or the membrane is flat. Below the critical temperature, the spins are polarized and the mechanical part of the system (string or membrane) is buckled. 

In the case of suspended graphene sheets, there is a large degree of arbitrariness in the assignation of numerical values to the spin-atom coupling constant and to the spin-flip rate\cite{bc12} and there is not yet unambiguous experimental evidence of a buckling transition at finite temperature. Instead, quantum theories point out that rippling due to the strong interplay between the free electron dynamics in the membrane and its mesoscopic structure,\cite{kim08,gaz09} already occurs at 0K.\cite{sgg11} Thus it is convenient to have a model exhibiting rippling at any temperature and whose parameters can be calibrated with experiments. In this paper we present such a model. 

We describe the position of carbon atoms by discrete elasticity equations previously used to model defect dynamics in graphene.\cite{CBJV08,car08,bon11} Neighboring carbon atoms in a graphene sheet attach to each other using three of their bonds. The fourth bond is not saturated and, similar to the effect of the electron beam, it may try to pull the atom upward or downward from the flat sheet configuration. This trend may be modeled by placing each atom in a double-well potential. This is reminiscent of the system of a soft spin coupled to elasticity used to study two-dimensional (2D) melting mediated by dislocations.\cite{CN96} Without the double-well potential, our numerical simulations of the time-dependent von Karman plate equations (discretized on the honeycomb lattice) show that a suspended sheet that is initially in a rippled state remains rippled whereas a flat sheet remains flat. In the absence of longitudinal stretching, the initial conditions of the plate equations select whether the graphene sheet remains flat or it acquires curvature or ripples. In experiments, rippling is always observed, so we have to look for a mechanism to produce it spontaneously. Theories including electron-phonon coupling indicate that, for infinitesimal stress, ripples are spontaneously formed even at 0K, \cite{sgg11} which introduces an extra length similar to that in our double-well potential. In addition, we add nonlinear friction and an appropriate white noise forcing that satisfies the fluctuation-dissipation theorem. The fundamental mode of the equations of motion can be projected on a Duffing oscillator with nonlinear friction and white noise forcing. It tuns out that Eichler et al have interpreted their measurements of graphene resonators by using such a Duffing model (without white noise forcing).\cite{eic11} In this latter work, measurements of carbon nanotube and graphene resonators are used to calculate the coefficients of the Duffing oscillator. Assuming that the minima of the potential are upward or downward a distance $w_0$ off-plane, we use their coefficients to fit the parameters in the equations of motion of our graphene model. The only free parameter in our model is $w_0$. While a more elaborated theory could derive a temperature and strain dependent $w_0$ from first principles (for example, considering an electron-phonon interaction)\cite{sgg11}, we have considered fixed values of $w_0$ from 0.5 to 1 nm for the sake of simplicity. Quantum theories produce lengths of this order.\cite{gaz09,sgg11} Numerical solutions of our model yield ripples without preferred orientation in agreement with experimental observations.\cite{mey07,ban09} 

The rest of the paper is as follows. The model we use is described in Section \ref{sec:2}. In Section \ref{sec:3}, we calculate numerical values of the coefficients used in our model by fitting it to one of the graphene resonators studied by Eichler et al.\cite{eic11} In Section \ref{sec:4}, we discretize the equations of motion on the honeycomb lattice, write them in primitive coordinates and periodize them, so that integer multiples of displacements on primitive directions leave the lattice unchanged. We then solve numerically the resulting stochastic equations of motion. We show that long-lived ripples with no preferred orientation emerge after a short transient even from random initial conditions. These ripples correspond to those observed in experiments.\cite{mey07,ban09} Lastly, Section \ref{sec:5} contains our conclusions.

\section{Model} 
\label{sec:2}

In the graphene sheet, carbon atoms have $\sigma$ bond orbitals constructed from $sp^2$ hybrid states oriented in the direction of the bond that accommodate three electrons per atom. The other electrons go to $p$ states oriented perpendicularly to the sheet. These orbitals bind covalently with neighboring atoms and form a narrow $\pi$ band that is half-filled. The presence of bending and ripples in graphene modifies its electronic structure.\cite{cas09} Out-of-plane convex or concave deformations of the sheet have in principle equal probability and transitions between these deformations are associated with the bending energy of the sheet. A simple way to model this situation is to consider that out-of-plane deformations are described by a double-well site potential that tries to set vertical deflections of the sheet, $w(x,y)$ to $\pm \tilde{w}_0$ and contributes the free energy:
\begin{eqnarray}
F_{DW}= \frac{\tilde{\varphi}}{4} \int \rho_2\!\left[1-\left(\!\frac{w(x,y)}{\tilde{w}_0}\right)^2\!\right]^2 dx\, dy, \label{eq3}
 \end{eqnarray}
where $\rho_2$ is the 2D mass density (mass per unit area) and $\tilde{\varphi}$ has units of velocity square. The elastic free energy of the graphene sheet in the continuum limit is that of a 2D membrane,\cite{nel02}
\begin{eqnarray}
&& F_g= \frac{1}{2}\int [\tilde{\kappa}(\nabla^2w)^2 + (\tilde{\lambda} u_{ii}^2 + 2\tilde{\mu} u_{ik}^2)]\, dx\, dy,  \label{eq1}\\
&& u_{ik}= \frac{1}{2}(\partial_{x_k}u_{i}+\partial_{x_i} u_k+\partial_{x_i}w\partial_{x_k}w),\, i,k=1,2,  \label{eq2}
\end{eqnarray}
where $(u_1,u_2)=(u(x,y),v(x,y))$, and $\tilde{\kappa}$, $\tilde{\lambda}$ and $\tilde{\mu}$ are the in-plane displacement vector, the bending stiffness (measured in units of energy) and the 2D Lam\'e coefficients of graphene (measured in units of force per unit length), respectively. $\nabla=(\partial_x,\partial_y)$ is the 2D gradient and $\nabla^2$ the 2D laplacian. In (\ref{eq2}) we have ignored the small in-plane nonlinear terms $\partial_{x_i}u\partial_{x_k}u+\partial_{x_i}v\partial_{x_k}v$.

From the total free energy $F=F_g+F_{DW}$, we obtain the equations of motion
 \begin{eqnarray}
 \rho_2\partial_t^2 u&=& \tilde{\lambda}\,\partial_x\left( \partial_x u + \partial_y v+\frac{|\nabla w|^2}{2}\right) + \tilde{\mu}\,\partial_x [2 \partial_x u + (\partial_xw)^2]\nonumber\\
&+& \tilde{\mu}\,\partial_y\left( \partial_y u + \partial_x v+ \partial_xw\partial_y w \right), \label{eq4}\\
 \rho_2\partial_t^2 v &=& \tilde{\lambda}\,\partial_y\left( \partial_x u + \partial_y v+\frac{|\nabla w|^2}{2}\right) +\tilde{\mu}\,\partial_y[2 \partial_y v + (\partial_yw)^2]\nonumber\\
&+&\tilde{\mu}\,\partial_x\left(\partial_y u+\partial_x v+\partial_xw\partial_y w \right),  \label{eq5}
\end{eqnarray}
\begin{eqnarray}
&&  \rho_2 \partial_t^2w = \tilde{P}\nabla^2w-\tilde{\kappa}\, (\nabla^2)^2w+\!\left(1-\frac{w^2}{\tilde{w}_0^2}\right)\!\frac{\tilde{\varphi}\rho_2}{\tilde{w}_0^2}w+\tilde{\lambda}\,\nabla\cdot\left[\!\left( \partial_x u + \partial_y v+\frac{|\nabla w|^2}{2}\right)\!\nabla w\right] \nonumber\\
&&\quad\quad\quad+ \tilde{\mu}\,\partial_x[2 \partial_x u\partial_xw +(\partial_y u + \partial_x v)\partial_yw+ |\nabla w|^2\partial_xw ] \nonumber\\
&&\quad\quad\quad+ \tilde{\mu}\,\partial_y[( \partial_y u + \partial_x v)\partial_xw+ 2\partial_y v\partial_y w+ |\nabla w|^2\partial_yw]\nonumber\\
&& \quad\quad\quad-(\tilde{\gamma}+\tilde{\eta}w^2)\partial_tw+\sqrt{2\tilde{\theta}(\tilde{\gamma}+\tilde{\eta}w^2)}\,\xi(x,y,t), \label{eq6}
\end{eqnarray}
\begin{eqnarray}
&&\langle\xi(x,y,t)\rangle=0, \quad \langle \xi(x,y,t)\xi(x,y,t)\rangle=\delta(x-x')\delta(y-y')\delta(t-t'),\label{eq7}
 \end{eqnarray}
where $\tilde{P}$ is the membrane stress, $\tilde{\theta}$ is the temperature measured in units of energy and $-(\tilde{\gamma}+\tilde{\eta}w^2)\partial_tw$ is a nonlinear friction force used by Eichler et al \cite{eic11} to interpret their experiments with  a forced damped graphene resonator. The intensity $\sqrt{2\tilde{\theta}(\tilde{\gamma}+\tilde{\eta}w^2)}$ of the white noise $\xi(t)$ is related to the friction by the fluctuation-dissipation theorem. All the parameters $\tilde{\lambda}$, $\tilde{\mu}$, $\rho_2$, $\tilde{\kappa}$, $\tilde{P}$, $\tilde{w}_0$, $\tilde{\varphi}$, $\tilde{\gamma}$, $\tilde{\eta}$ and $\tilde{\theta}$ are positive. 

\section{Parameter identification}
\label{sec:3}
To identify the parameters $\tilde{\lambda}$, $\tilde{\mu}$, $\rho_2$, $\tilde{\kappa}$, $\tilde{P}$, $\tilde{w}_0$, $\tilde{\varphi}$, $\tilde{\gamma}$, $\tilde{\eta}$ and $\tilde{\theta}$, we shall consider the graphene resonator corresponding to Figure 3 in Ref.Ê\onlinecite{eic11}. The temperature is 4K, so that we use $\tilde{\lambda}=3.25$ eV/\AA$^2$ and $\tilde{\mu}=9.44$ eV/\AA$^2$ corresponding to 0K in Table 1 of Ref. \onlinecite{zak09}. The graphene resonator has length $L=1.7\,\mu$m, width $W=120$ nm and is pinned at $x=0$ and $x=L$. The sides at $y=0$ and $y=W$ are free.\cite{eic11} Let us assume that $u=v=0$, ignore bending and noise, and assume that $w$ is in the lowest possible eigenstate:
\begin{eqnarray}
w(x,y,t)=\Omega(t)\,\sin\frac{\pi x}{L}.\label{eq8}
\end{eqnarray}
We now insert Eq.\ (\ref{eq8}) in (\ref{eq6}), multiply the result by $\sin(\pi x/L)$, integrate and divide by $\int_0^L\sin^2(\pi x/L)dx=L/2$. We obtain 
\begin{eqnarray}
\rho_2\frac{d^2\Omega}{dt^2}=-\frac{\tilde{P}\pi^2}{L^2}\Omega+\frac{\tilde{\varphi}\rho_2}{\tilde{w}_0^2}\Omega-\frac{3\tilde{\varphi}\rho_2}{4\tilde{w}_0^4}\Omega^3-\frac{3\pi^4}{8L^4}(\tilde{\lambda}+2\tilde{\mu})\Omega^3-\left(\tilde{\gamma}+\frac{3\tilde{\eta}}{4}\Omega^2\right)\!\frac{d\Omega}{dt}. \label{eq9}
\end{eqnarray}
The stress $\tilde{P}$ in the sheet is due to the inherent stretching of a stressed planar membrane, $\tilde{P}_0=E_2\varepsilon_0$ [where $\varepsilon_0$ is the homogeneous strain and $E_2=4\tilde{\mu}(\tilde{\lambda}+\tilde{\mu})/(\tilde{\lambda}+2\tilde{\mu})$ is the 2D Young's modulus], and an additional stress $\Delta\tilde{P}$ due to its bending. The latter is proportional to the variation of the membrane area:
\begin{eqnarray}
&& \frac{\Delta S}{S}=\int_0^L\int_0^W\frac{\sqrt{1+|\nabla w|^2}-1}{LW}\, dxdy\approx \frac{1}{2LW}\int_0^L\int_0^W|\nabla w|^2 dxdy=\frac{\pi^2}{4L^2}\Omega^2,  \label{eq10}
\end{eqnarray}
in which Eq.\ (\ref{eq8}) has been used. Thus the total stress $\tilde{P}$ is
\begin{eqnarray}
\tilde{P}=E_2\!\left(\varepsilon_0+ \frac{\pi^2}{4L^2}\Omega^2\right)\!,\quad E_2=\frac{4\tilde{\mu}(\tilde{\lambda}+\tilde{\mu})}{\tilde{\lambda}+2 \tilde{\mu}}, \label{eq11}
 \end{eqnarray}
which inserted in Eq.\ (\ref{eq9}) produces
\begin{eqnarray}
\frac{d^2\Omega}{dt^2}&=&-\left(\frac{\tilde{P}_0\pi^2}{\rho_2L^2}-\frac{\tilde{\varphi}}{\tilde{w}_0^2}\right)\!\Omega-\left(\frac{3\tilde{\varphi}}{4\tilde{w}_0^4}+\frac{3\pi^4}{8\rho_2L^4}(\tilde{\lambda}+2\tilde{\mu})+E_2\frac{\pi^4}{4\rho_2L^4}\right)\!\Omega^3\nonumber\\
&-&\left(\frac{\tilde{\gamma}}{\rho_2}+\frac{3\tilde{\eta}}{4\rho_2}\Omega^2\right)\!\frac{d\Omega}{dt}. \label{eq12}
 \end{eqnarray}
This should be identified with the nonlinearly damped Duffing oscillator used by Eichler et \cite{eic11} as a model of their graphene resonator:
\begin{eqnarray}
\ddot{x}=-\frac{k}{m}x-\frac{\alpha}{m}x^3-\frac{\gamma^*+\eta^* x^2}{m}\dot{x}, \label{eq13}
\end{eqnarray}
where $k/m=\pi^2T_0/(mL)$ is the resonant frequency squared, $m=3.9\times 10^{-19}$kg is the mass of the sheet, $T_0=110$ nN is the intrinsic tension, $\alpha=1.4\times 10^{16}$ kg m$^{-2}$ s$^{-2}$, $\gamma^*=8.7\times 10^{-14}$ kg/s, and $\eta^*=1.5\times 10^7$ kg m$^{-2}$s$^{-1}$. We find 
\begin{eqnarray}
\frac{\tilde{\varphi}}{\tilde{w}_0^4}=\frac{4\alpha}{3m} -\frac{\pi^4(\tilde{\lambda}+2\tilde{\mu})}{\rho_2L^4}\! \left(\frac{1}{2}+ \frac{4\tilde{\mu}(\tilde{\lambda}+\tilde{\mu})}{3(\tilde{\lambda}+2\tilde{\mu})^2}\right)\!= 4.3142\times 10^{34}\,\mbox{m$^{-2}$s$^{-2}$.}\label{eq14}
 \end{eqnarray}
Setting $\tilde{w}_0=5$ \AA, we obtain $\tilde{\varphi}=2.6964\times 10^{-3}$ m$^{2}$s$^{-2}$, an intrinsic strain $\varepsilon_0=0.0010067$ and stress $\tilde{P}_0=0.3492$ kg s$^{-2}$. The friction coefficients are $\tilde{\gamma}/\rho_2= 2.2308\times 10^5$ s$^{-1}$ and $\tilde{\eta}/\rho_2=5.1282\times 10^{25}$ m$^{-2}$s$^{-1}$.

\section{Numerical results } 
\label{sec:4}
\begin{figure}
\begin{center}
\includegraphics[width=8cm]{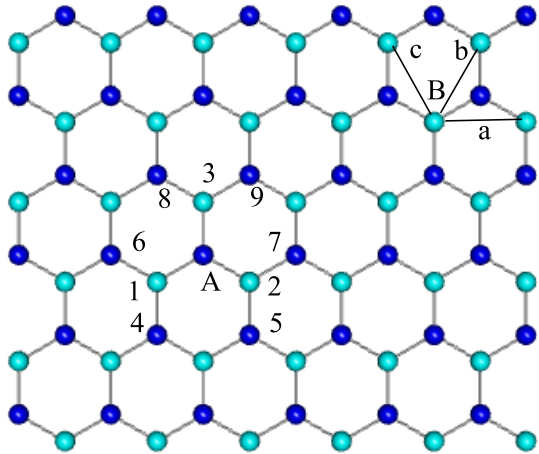}
\caption{(Color online) Neighbors of a given atom $A$ in sublattice 1 (dark blue). The primitive vectors are ${\bf a}$, ${\bf b}$ and ${\bf c}$ that connect atom $B$ to three different next nearest neighbors in its same sublattice. }
\label{fig1}
\end{center}
\end{figure}

Following the same procedure as in References \onlinecite{car08,CBJV08,bon11}, we discretize the equations of motion on the honeycomb lattice of graphene.  Atoms belonging to the two sublattices comprising the lattice are distinguished in Figure \ref{fig1}with light and dark colors, although they are all identical carbon atoms. The three nearest and six next nearest neighbors of the atom $A$ with coordinates $(x,y)$ in Fig. \ref{fig1} are 
\begin{eqnarray}
&& n_{1}=\left(x-\frac{a}{ 2},y-\frac{a}{ 2\sqrt{3}}\right)\!\!, \, n_{2}=\left(x+\frac{a}{ 2}, y-\frac{a}{ 2\sqrt{3}}\right)\!\!, \nonumber\\ 
&& n_{3}=\left(x,y+\frac{a}{\sqrt{3}}\right)\!,\quad n_{4}=\left(x-\frac{a}{ 2},y-\frac{a\sqrt{3}}{ 2}\right)\!,\nonumber\\
&& n_{5}=\left(x+\frac{a}{ 2},y-\frac{a\sqrt{3}}{ 2}\right)\!, \quad n_{6}=(x-a,y), \nonumber\\ 
&& n_{7}=(x+a,y),\quad n_{8}=\left(x-\frac{a}{ 2},y+\frac{a\sqrt{3}}{ 2}\right)\!,\nonumber\\
&& n_{9}=\left(x+\frac{a}{ 2},y+\frac{a\sqrt{3}}{ 2}\right)\!. \label{eq15}
\end{eqnarray}
To discretize the equations of motion using the honeycomb lattice, we define the following operators acting on functions of the node $A$ coordinates $(x,y)$,
\begin{eqnarray}
Tu &=& [u(n_{1})-u(A)] + [u(n_{2})-u(A)] \nonumber\\
&+& [u(n_{3})-u(A)]\sim (\partial_x^2 u + \partial_y^2 u)\,\frac{a^2}{4},\label{eq16}\\
Hu &=& [u(n_{6})-u(A)]+[u(n_{7})-u(A)]\sim a^2\partial_x^2 u,\,\,  \label{eq17} \\
D_{1}u &=& [u(n_{4})-u(A)] + [u(n_{9})-u(A)]  \label{eq18}\\
&\sim& \left(\frac{1}{ 4}\, \partial_x^2 u +\frac{\sqrt{3}}{ 2}\,\partial_x\partial_yu + \frac{3}{ 4}\, \partial_y^2 u\right)a^2,\nonumber\\
D_{2}u &=& [u(n_{5})-u(A)] + [u(n_{8})-u(A)]  \label{eq19}\\
&\sim&\left(\frac{1}{ 4}\, \partial_x^2 u -\frac{\sqrt{3}}{ 2}\,\partial_x\partial_yu + \frac{3}{ 4}\, \partial_y^2 u\right)a^2,  \nonumber\\
\Delta_hu &=& u(n_{7})-u(A)\sim (\partial_xu)\, a, \label{eq20}\\
\Delta_v u &=& u(n_{3})-u(A)\sim (\partial_yu)\, \frac{a}{\sqrt{3}}, \label{eq21}\\
Bw&=& [Tw(n_1) - Tw(A)] + [Tw(n_2) - Tw(A)] \nonumber\\
&+& [Tw(n_3) - Tw(A)]\sim \frac{a^4}{16} (\partial_x^2+\partial_y^2)^2w.\label{eq22}
\end{eqnarray}
with similar definitions for points $B$ in sublattice 2. Taylor expansions of these finite difference combinations about $(x,y)$ yield the partial derivative expressions written above as $a\to 0$. We now replace the derivatives in (\ref{eq4}), (\ref{eq5}) and (\ref{eq6}), $\partial_x^2 u$, $\partial_y^2 u$, $\partial_x\partial_yu$, $\partial_x w$, $\partial_yw$ and $(\nabla^2)^2w$ by the operators $Hu/a^2$, $(4T-H)u/a^2$, $(D_{1}-D_{2})u/(\sqrt{3}a^2)$, $\Delta_hw/a$, $\sqrt{3}\Delta_vw/a$ and $16 Bw/a^4$, respectively, with similar substitutions for the derivatives of $v$ and $w$. The resulting equations at each point of the lattice are
\begin{eqnarray}
\rho_2 a^2 \partial_t^2 u &=& 4\tilde{\mu}\, Tu + (\tilde{\lambda}+\tilde{\mu})\, Hu +\frac{\tilde{\lambda}+\tilde{\mu} }{\sqrt{3}}\, (D_{1} - D_{2})v  \nonumber\\
&+&\frac{\tilde{\lambda}+\tilde{\mu}}{a}\, [\Delta_hw\, Hw+\Delta_vw\, (D_1-D_2)w]+ \frac{4\tilde{\mu}}{a}\, \Delta_hw\, Tw, \label{eq23}\\
\rho_2 a^2 \partial_t^2 v &=& 4 (\tilde{\lambda}+2\tilde{\mu} )\, Tv +\frac{\tilde{\lambda}+\tilde{\mu}}{\sqrt{3}}\, (D_{1} - D_{2})u \nonumber\\
&-& (\tilde{\lambda}+\tilde{\mu} )H v+ \frac{4\sqrt{3}}{a} (\tilde{\lambda}+2\tilde{\mu})\Delta_vw\, Tw\nonumber\\
&+&\frac{\tilde{\lambda}+\tilde{\mu}}{a\sqrt{3}}\, [\Delta_hw\, (D_1-D_2)w-3\Delta_vw\, Hw], \label{eq24}
\end{eqnarray}
\begin{eqnarray}
\rho_2 a^2\partial_t^2w &=& \frac{\tilde{\lambda}+2\tilde{\mu}}{a}\left\{\left[Hu+\frac{2\Delta_vw}{a}(D_1-D_2)w +\frac{\Delta_hw}{a}Hw\right]\Delta_hw \right.\nonumber\\
&+& \left[ \sqrt{3}(4T-H)v
\left.+\frac{3\Delta_vw}{a}(4T-H)w\right]\Delta_vw\right\}\nonumber\\
&+& \frac{\tilde{\lambda}+\tilde{\mu}}{a}(D_1-D_2)u\Delta_vw+\frac{\tilde{\lambda}\Delta_hw}{\sqrt{3}a}(D_1-D_2)v\nonumber\\
&+&\frac{\tilde{\mu}\Delta_vw}{a}[\sqrt{3}(4T-H)u+\sqrt{3}Hv+(D_1-D_2)v]+\frac{2\Delta_hu}{a}(2\tilde{\lambda} T+\tilde{\mu} H)w\nonumber\\
&+&\frac{2\sqrt{3}\Delta_vv}{a}[2\tilde{\lambda} Tw+\tilde{\mu} (4T-H)w]+\frac{2\tilde{\mu}}{a}\left(\Delta_v u+\frac{\Delta_h v}{\sqrt{3}}\right)(D_1-D_2)w\nonumber\\
&+&\frac{(\Delta_hw)^2+(\Delta_vw)^2}{a^2}(2\tilde{\lambda} T+\tilde{\mu} H)w+\frac{4\tilde{\mu}}{a}Tw(4T-H)w\nonumber\\
&+& 4\tilde{P}\, Tw-\frac{16\tilde{\kappa}}{a^2} Bw+\!\left(1-\frac{w^2}{\tilde{w}_0^2}\right)\!\frac{\tilde{\varphi}\rho_2a^2}{\tilde{w}_0^2}w-a^2(\tilde{\gamma}+\tilde{\eta}w^2)\partial_tw\nonumber\\
&+&\sqrt{2a^2\theta(\tilde{\gamma}+\tilde{\eta}w^2)}\, a\xi(x,y,t).\label{eq25}
\end{eqnarray}
Here
\begin{eqnarray}
\tilde{P}=\frac{4\tilde{\mu}(\tilde{\lambda}+\tilde{\mu})}{\tilde{\lambda}+2\tilde{\mu}}\!\left(\varepsilon_0+\frac{1}{2Na^2}\sum_{x,y}[(\Delta_hw)^2+9(\Delta_vw)^2] \right)\!, \label{eq26}
\end{eqnarray}
is the stress in the membrane (see section \ref{sec:3}) and $N=\sum_{x,y} 1$ is the total number of atoms in the graphene sheet. In the limit $a\to 0$, the continuous delta function $\delta(x-x')$ becomes $\delta_{xx'}/a$, where $\delta_{xx'}=1$ if $x=x'$ and 0 otherwise. Then the discretized white noise in Eq.\ (\ref{eq25}), $a\xi(x,y,t)$, has zero mean and correlation 
\begin{eqnarray}
\langle a\xi(x,y,t)\, a\xi(x',y',t')\rangle=\delta_{xx'}\delta_{yy'}\delta(t-t'). \label{eq27}
\end{eqnarray}
Possible defects inserted in the graphene sheet are the cores of dislocations. To account for them, we have to write these equations of motion in primitive coordinates and periodize all difference operators appearing in them along  primitive directions. The resulting equations of motion become those in Ref. \onlinecite{car08} for $w=0$. \textit{If there are not defects in the graphene sheet, the equations of motion (\ref{eq23})-(\ref{eq25}) or their periodized version produce the same results}. 

\begin{table}[ht]
\begin{center}\begin{tabular}{ccccc} \hline
 $u$, $v$, $w$, $\tilde{w}_0$ &  $t$& $\xi$ &$\tilde{\theta}$ & $\tilde{P}$   \\
$a$ &  $\sqrt{\frac{\rho_{2}a^2}{\tilde{\lambda}+2\tilde{\mu}}}$& $\left(\frac{\tilde{\lambda}+2\tilde{\mu}}{\rho_2a^6}\right)^{1/4}$ & $(\tilde{\lambda}+2\tilde{\mu})\, a^2$ &$\tilde{\lambda}+2\tilde{\mu}$  \\ \hline
\AA & $10^{-14}$ s& m$^{-1}$s$^{-1/2}$ & eV &eV/$\AA^2$\\
1.42$\sqrt 3$&  1.11109& $3.85723\times 10^{16}$ & 133.8688 &22.13\\
 \hline
\end{tabular}
\end{center}
\caption{Units used to nondimensionalize equations. } \label{t1}
\end{table}

It is convenient to write the previous equations of motion in nondimensional units. Using Table \ref{t1}, we obtain the following nondimentional equations of motion 
\begin{eqnarray}
\partial_t^2 u &=& 4\mu\, Tu + (\lambda+\mu)\left( Hu +\frac{(D_{1} - D_{2})v }{\sqrt{3}}\right)   \nonumber\\
&+&(\lambda+\mu)\, [\Delta_hw\, Hw+\Delta_vw\, (D_1-D_2)w]+ 4\mu\, \Delta_hw\, Tw, \label{eq28}\\
\partial_t^2 v &=& 4 Tv +(\lambda+\mu) \left(\frac{(D_{1} - D_{2})u}{\sqrt{3}}- H v\right)\!+ 4\sqrt{3}\,\Delta_vw\, Tw\nonumber\\
&+&\frac{\lambda+\mu}{\sqrt{3}}\, [\Delta_hw\, (D_1-D_2)w-3\Delta_vw\, Hw], \label{eq29}
\end{eqnarray}
\begin{eqnarray}
\partial_t^2w &=& \left[Hu+2\Delta_vw\, (D_1-D_2)w +\Delta_hw\, Hw\right]\Delta_hw \nonumber\\
&+& \left[ \sqrt{3}(4T-H)v+ 3\Delta_vw\, (4T-H)w\right]\Delta_vw \nonumber\\
&+& (\lambda+\mu)(D_1-D_2)u\,\Delta_vw+\frac{\lambda}{\sqrt{3}}\, \Delta_hw\, (D_1-D_2)v\nonumber\\
&+&\mu\,\Delta_vw\, [\sqrt{3}(4T-H)u+\sqrt{3}Hv+(D_1-D_2)v]+2\Delta_hu\, (2\lambda T+\mu H)w\nonumber\\
&+&2\sqrt{3}\Delta_vv\, [2\lambda T+\mu (4T-H)]w +2\mu\!\left(\Delta_v u+\frac{\Delta_h v}{\sqrt{3}}\right)\! (D_1-D_2)w\nonumber\\
&+&[(\Delta_hw)^2+(\Delta_vw)^2](2\lambda T+\mu H)w+4\mu\, Tw\, (4T-H)w+4P\, Tw\nonumber\\
&-&\kappa Bw+\!\varphi\left(1-\frac{w^2}{w_0^2}\right)\!\frac{w}{w_0^2}-(\gamma+\eta w^2)\,\partial_tw
+\sqrt{2\theta(\gamma+\eta w^2)}\, \Xi(x,y,t),\label{eq30}
\end{eqnarray}
where $u$, $v$, $w$ and $t$ are now nondimensional and 
\begin{eqnarray}
&& P=\frac{4\tilde{\mu}(\tilde{\lambda}+\tilde{\mu})}{(\tilde{\lambda}+2\tilde{\mu})^2}\!\left(\varepsilon_0+\frac{1}{2N}\sum_{x,y}[(\Delta_hw)^2+9(\Delta_vw)^2] \right)\!, \label{eq31}\\
&&\lambda=\frac{\tilde{\lambda}}{\tilde{\lambda}+2\tilde{\mu}},\quad\mu=\frac{\tilde{\mu}}{\tilde{\lambda}+2\tilde{\mu}}, \label{eq32}\\
&&\langle\Xi(x,y,t)\rangle=0,\quad\langle \Xi(x,y,t)\,\Xi(x',y',t')\rangle=\delta_{xx'}\delta_{yy'}\delta(t-t'). \label{eq33}
\end{eqnarray}
The dimensionless parameters appearing in these equations are listed in Table \ref{t2}. We have used the values of the 2D Lam\'e moduli at 0K given in Ref.~\onlinecite{zak09}: $\tilde{\lambda}+2\tilde{\mu}=22.13$ eV/\AA$^2$, $\tilde{\mu}=9.44 $ eV/\AA$^2$. The bending rigidity $\tilde{\kappa}=0.8$ eV has been extrapolated from Fig. 7 of Ref. \onlinecite{zak10}. The 2D mass density is $\rho_2=7.236\times 10^{-7}$ kg/m$^2$, and the parameters identified in Section \ref{sec:3} are $\tilde{\varphi}/\tilde{w}_0^4=4.3142\times 10^{34}$ m$^{-2}$s$^{-2}$ and, for $\tilde{w}_0=5$ \AA, we have $\varepsilon_0=0.0010067$,  $P_0= 0.3492$ kg s$^{-2}$, $\tilde{\gamma}/\rho_2= 2.2308 \times 10^5$ s$^{-1}$ and $\tilde{\eta}/\rho_2=5.1282\times 10^{25}$ m$^{-2}$s$^{-1}$. In our simulations, we have considered shorter samples than in Eichler et al's experiments for the sake of computational expediency. To keep the same frequency, we need to rescale $\varphi$ as $L^{-4}$, and both $\eta$ and $\gamma$ as $1/L$. For a $80\times 80$ lattice, this yields the parameters in the second row of Table \ref{t2}. 

\begin{table}[ht]
\begin{center}\begin{tabular}{ccccccc}
 \hline
$w_0$&$\theta$  &$\kappa$&$\gamma$& $\eta$ & $\varphi$& $\varepsilon_0$ \\
$\frac{\tilde{w}_0}{a}$&$\frac{\tilde{\theta}}{(\tilde{\lambda}+2\tilde{\mu})a^2}$ &$\frac{16\tilde{\kappa}}{(\tilde{\lambda}+2\tilde{\mu})a^2}$& $\frac{\tilde{\gamma} a}{\sqrt{\rho_2(\tilde{\lambda}+2\tilde{\mu})}}$& $\frac{\tilde{\eta}a^3}{\sqrt{\rho_2(\tilde{\lambda}+2\tilde{\mu})}}$&$\frac{\rho_2\tilde{\varphi}}{\tilde{\lambda}+2\tilde{\mu}}$&\\ \hline
2.03292&$2.57484\times 10^{-6}$ & $0.095616$&0.00342536& 0.0476338 & $5.5028\times 10^{-12}$&0.0010067\\
2.03292&$2.57484\times 10^{-6}$ & $0.095616$&0.295949& 4.11553 & $3.06636\times 10^{-4}$&0.0010067\\
 \hline
\end{tabular}
\end{center}
\caption{Dimensionless parameters for Fig.\ 3 of Ref.~\onlinecite{eic11} with $\tilde{w}_0=5$ \AA. The numbers in the last row correspond to a shorter computational lattice with $L=80a$.} 
\label{t2}
\end{table}

\begin{figure}
\begin{center}
\includegraphics[width=14cm]{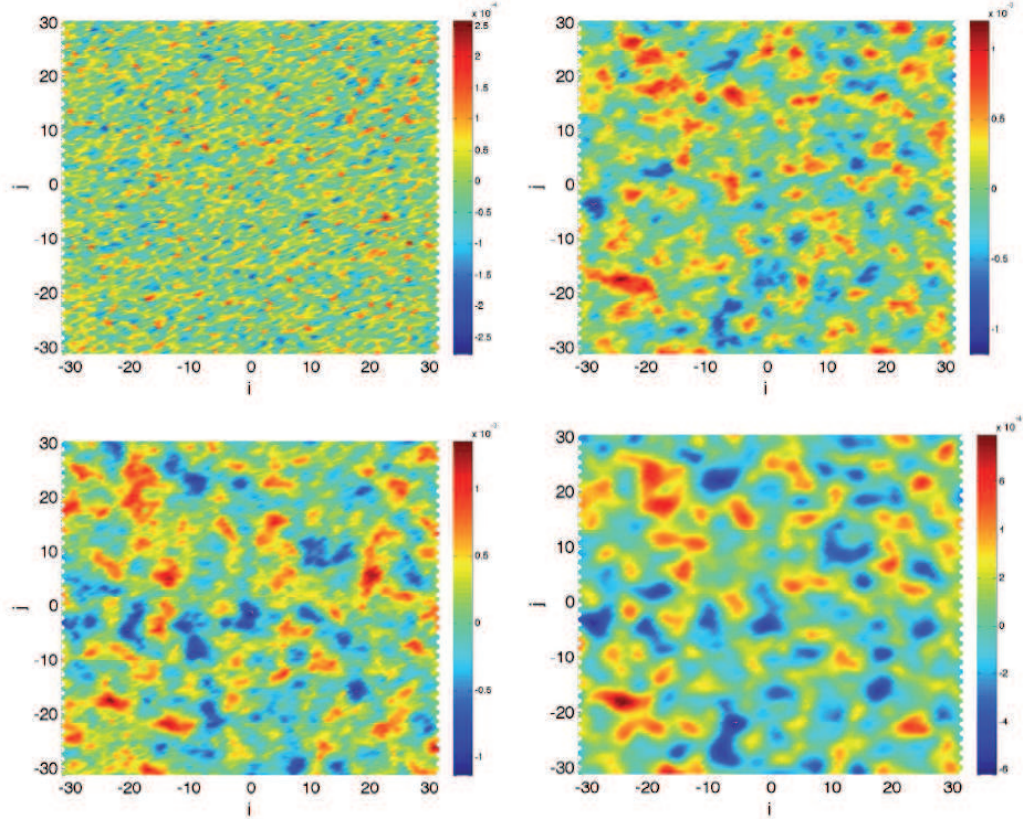}
\caption{(Color online) Ripples formed in a suspended $80\times 80$-honeycomb graphene sheet (with 12,800 atoms). (a) Density plot of the off-plane displacement as a function of in-plane coordinates at $t=1$. (b) The same at $t=50$, and (c) $t=100$. (d) Average density plot over all times from $t=1$ to 100 with unit time step. Parameter values as in the second row of Table \ref{t2}. }
\label{fig2}
\end{center}
\end{figure}

\begin{figure}
\begin{center}
\includegraphics[width=10.5cm,angle=0]{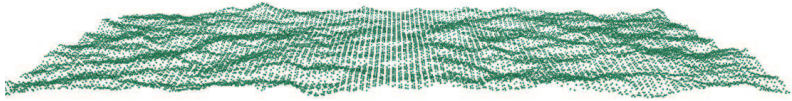}\\
\caption{(Color online) Ripples in the central part of a $160\times 160$ graphene lattice (with 51,200 atoms).}
\label{fig3}
\end{center}
\end{figure}

\begin{figure}
\begin{center}
\includegraphics[width=14cm]{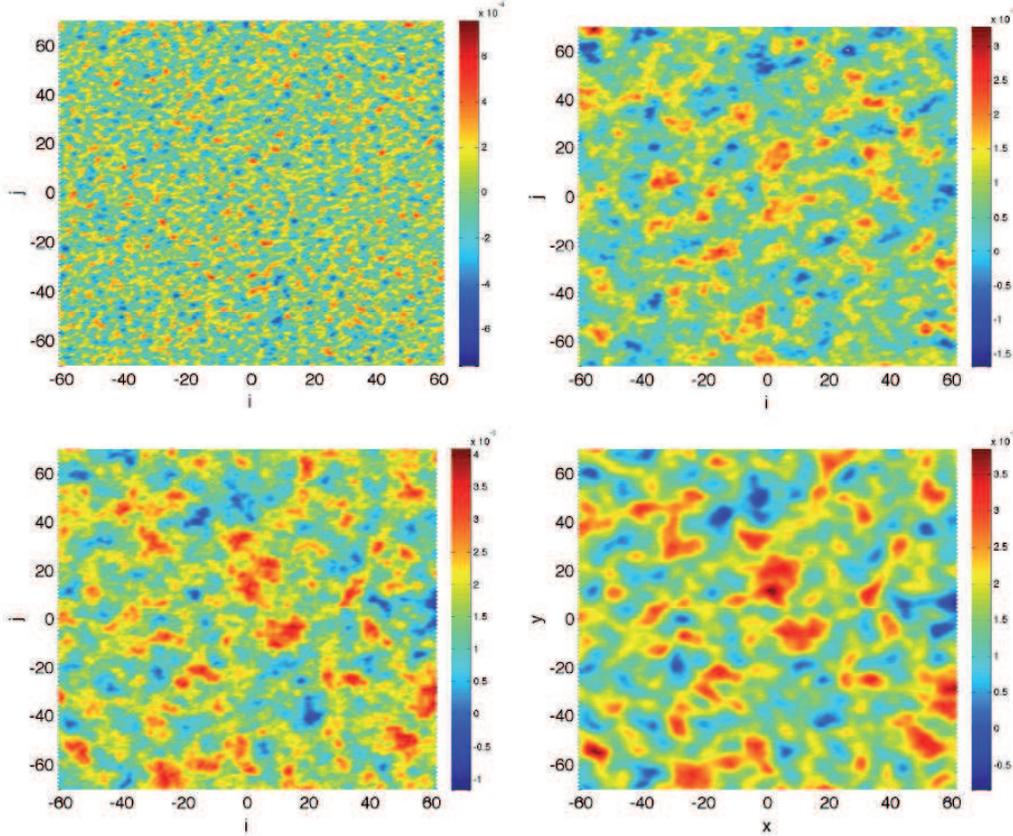}
\caption{(Color online) Same as in Fig. \ref{fig1} for a $200\times 200$ honeycomb lattice with $w_0=1$ nm (180,000 atoms). Times are (a) $t=10$, (b) 250, (c) 510. (d) Average density plot from $t=410$ to 510, using a $t=10$ time step. Parameters are as in Table \ref{t1} except for the rescaled $\gamma= 0.11838$, $\eta=1.64621$, $\varphi= 1.25598 \times 10^{-4}$.}
\label{fig4}
\end{center}
\end{figure}

The model equations produce spontaneous ripple generation in which ripples have no preferred orientation and their size is comparable with that observed in experiments. We have considered that the graphene sheet has a strain similar to that in Ref. \onlinecite{eic11}. This subjacent strain flattens the ripples as compared to those in an unstrained sheet. Initially the graphene sample is flat. As shown in Fig. \ref{fig2}(a), random forcing pulls atoms off-plane and randomly oriented ripples that are very small in size and height form. After a short transient stage, the ripples have reached a quasi-stationary state in which they are domains of atoms displaced variable distances either upward or downward the horizontal plane, cf. Fig. \ref{fig2}(b) and (c) and Fig. \ref{fig3} (the latter, for a $160\times 160$ lattice). Ripple domains change slowly in height and size, essentially due to displacements of the atoms in their boundaries. This is seen when plotting the time averaged density plot of Fig. \ref{fig2}(d). More clearly, the averages between more advanced times (between 300 and 400) do not change much from the instantaneous time resolved density plots, as shown in Fig. \ref{fig4} for a larger $300\times 300$ graphene lattice.  

In the quasi-stationary state, ripple size and height increase with sample size and with $w_0$. For instance, in a $80\times 80$ lattice, ripples are 2.5 nm long and 0.1 pm high for $w_0=5$\AA. Increasing $w_0$ a factor 10 does not change the ripple height that much but there appear long connected domains with atoms either up or down the horizontal plane. For longer $160\times 160$ samples with $w_0=1$nm as in Fig. \ref{fig3}, the ripple length has increased to about 10 nm with height of about 0.5 pm.  Increasing sample size to $300\times 300$ results in longer ripples (15 nm) and height about 1 pm, as shown in Fig. \ref{fig4}. As graphene sheets are suspended on holes that have 1 $\mu$m diameter, the effects of boundaries are felt comparatively less by atoms far from them and the ripple heights increase correspondingly to several \AA ngstroms.\cite{mey07,ban09}  

\section{Conclusions} 
\label{sec:5}
We have proposed a mechanism to spontaneously produce ripples in a suspended graphene sheet. Besides the membrane free energy, carbon atoms are placed in a double-well potential at each lattice site. Thus they experience a force that tries to displace them vertically off-plane, either up or down with equal probability. Inspired by experiments with graphene resonators, we assume that the atoms undergo a nonlinear friction force and are subject to the corresponding multiplicative white noise force satisfying the fluctuation-dissipation theorem. The resulting time scale of friction is much larger than the vibration periods of the membrane. Both the site double-well potential and the site nonlinear friction should be considered as simple mechanisms that provide a nonlinearly damped Duffing oscillator equation for the amplitude of the leading membrane normal mode. Extra soft-spin fields\cite{CN96} or a viscous component of the stress tensor plus a white noise source as in fluctuating hydrodynamics\cite{ll6} may prove better options and yet yield a similar Duffing oscillator equation. We have solved numerically the equations of the model using parameters obtained from ab-initio calculations and from experiments in the literature on graphene resonators. Even from initial conditions corresponding to a flat membrane, the nonlinear force created by the double-well potential and the white noise forcing produce stable rippling after a short transient. The ripples are formed by domains all whose atoms are displaced upward or downward off plane. Once they have acquired a sufficient size, of the order of experimentally observed ripples, the domains vary slowly by annexing or losing atoms in their periphery. In this scenario, randomly oriented nanometer-sized ripples with no preferred direction appear as long-lived metastable states for any temperature. 

\acknowledgments
This work has been supported by the Spanish Ministerio de Econom\'\i a y Competitividad grants FIS2011-28838-C02-01, FIS2011-28838-C02-02 and FIS2010-22438-E (Spanish National Network Physics of Out-of-Equilibrium Systems). The authors thank M.P. Brenner for hospitality during a stay at Harvard University, financed by  Fundaci\'on Caja Madrid mobility grants.

\end{document}